# Comparison of Fe and Ni opacity calculations for a better understanding of pulsating stellar envelops

D. Gilles [a,*], S. Turck-Chièze [a], G. Loisel [a], L. Piau [a], J.-E. Ducret [a], M. Poirier [b], T. Blenski [b], F. Thais [b], C. Blancard [c], P. Cossé [c], G. Faussurier [c], F. Gilleron [c], J.C. Pain [c], Q. Porcherot [c], J.A. Guzik [d], D.P. Kilcrease [d], N.H. Magee [d], J. Harris [e], M. Busquet [f], F. Delahaye [g], C.J. Zeippen [g], S. Bastiani-Ceccotti [h]

[a] CEA/IRFU/SAp, F-91191 Gif-sur-Yvette France,

[b] CEA/IRAMIS/SPAM, F-91191 Gif-sur-Yvette France,

[c] CEA/DAM/DIF, F-91297 Arpajon, France,

[d] Theoretical Division, LANL, Los Alamos NM 87545, USA,

[e] AWE Reading, Berkshire, RG7 4PR, UK,

[f] ARTEP, Ellicott City MD 21042,

[g] LERMA, Observatoire de Paris, France,

[h] LULI, Ecole Polytechnique, CNRS, CEA, UPMC, F-91128 Palaiseau Cedex, France.

[*] Corresponding Author.
E-mail address: dominique.gilles@cea.fr,
CEA/IRFU/SAp, F-91191 Gif-sur-Yvette France, Phone: (33) 1 69 08 56 07, Fax: (33) 1 69 08 65 77

## Abstract:

Opacity is an important ingredient of the evolution of stars. The calculation of opacity coefficients is complicated by the fact that the plasma contains partially ionized heavy ions that contribute to opacity dominated by H and He. Up to now, the astrophysical community has greatly benefited from the work of the contributions of Los Alamos [1], Livermore [2] and the Opacity Project (OP) [3]. However unexplained differences of up to 50% in the radiative forces and Rosseland mean values for Fe have been noticed for conditions corresponding to stellar envelopes. Such uncertainty has a real impact on the understanding of pulsating stellar envelopes, on the excitation of modes, and on the identification of the mode frequencies. Temperature and density conditions equivalent to those found in stars can now be produced in laboratory experiments for various atomic species. Recently the photo-absorption spectra of nickel and iron plasmas have been measured during the LULI 2010 campaign, for temperatures between 15 and 40 eV and densities of ~3 mg/cm$^3$. A large theoretical collaboration, the "OPAC", has been formed to prepare these experiments. We present here the set of opacity calculations performed by eight different groups for conditions relevant to the LULI 2010 experiment and to astrophysical stellar envelope conditions.

# 1. Introduction

One fundamental ingredient for the evolution of stars is the opacity. The calculation of the stellar plasma opacity coefficients is complex due to the composition of these plasmas, generally a predominantly H /He mixture with a low concentration of heavier ions. The astrophysical community has primarily used the opacities from Los Alamos [1], Livermore [2], and the Opacity Project (OP) [3]. Only the OP data set contain the corresponding photon spectra. However, up to 50% differences in the radiative forces [4] have been noticed and even more in the Rosseland mean value for Cr, Mn and Ni that are still unexplained [5]. Such uncertainty has a real impact on the understanding of the excitation modes and on the identification of the mode frequencies [6-9]. Very recently Lenz et al. [10], Salmon et al. [11] and Miglio et al. [12] have shown the sensitivity of hybrid stars, e.g., γPeg and 44 Tau that have BCEP and SPB modes, to whether one uses OPAL and OP opacities. Comparisons with seismic observations indicate that neither OP and OPAL can reproduce all the information on SPB stars [13].

In the last decade opacity measurements on iron and nickel relevant to astrophysical cases have been performed [14-19], but the discrepancies between various models still remain. Fortunately high-energy laser facilities can produce plasmas with ionized populations as the stellar conditions of temperature and density of the iron peak for Cr, Fe, Ni, Cu and Ge [20]. More recently the photo-absorption spectra of nickel and iron plasmas have been measured during the LULI 2010 campaign, for temperatures between 15 and 40 eV and densities on the order of 2-3 mg/cm$^3$ [21-22]. The experiment and some initial results have been reported [22, 23].

We have begun a theoretical collaboration to analyze the experiment, motivated by the fact that many groups can now provide frequency-dependent opacity for the conditions to be achieved. In this paper we present the panel of theoretical calculations performed by eight groups, for conditions relevant to the LULI 2010 experiment and useful for the astrophysical envelope conditions. Models used were statistical (SCO [24], Cassandra [25], STA [26]), detailed (OPAS [27], LEDCOP [28], OP [3], FLYCHK [29]) or hybrid (SCO-RCG [30]). Atomic structure calculations performed using HULLAC [31] have also been done for iron and nickel. Thus, the original experimental motivation provides the opportunity to compare calculations of the frequency-dependent opacity for astrophysical conditions and this could also serve the opacity community.

In Sec. 2 we define the mean thermodynamic conditions of the chosen cases. In Sec. 3 we illustrate the theoretical predictions of the eight codes for the photon energy spectra of iron and nickel and we present some comparisons. Not surprisingly, codes present better agreement among them for the higher temperature (38.5eV) than at 15.3eV. We also compare Planck and Rosseland means for iron and nickel before we conclude.

# 2. From theoretical estimates to laboratory experiments

It is well established that β Cephei stars pulsate by the κ mechanism due to the iron group. A detailed discussion is provided in Ref. [5] on the consequences of differences found in the available sources of opacity for astrophysics.

At relatively low temperature, the calculations are very complex and performing laboratory experiments are useful to make progress on the relevant hypothesis.

To understand differences in Rosseland mean values, one needs to understand the frequency-dependent opacity model, i.e., the ingredients –ionic population distributions, configuration set included, averaging scheme, line shapes– and approximations. In the OP, frequency-dependent opacities are tabulated on line, together with their Rosseland and Planck means. Unfortunately, among the available OPAL data, only Rosseland and Planck means calculations can be found on the web and not the corresponding frequency-dependent opacities which are important for the interpretation of the asteroseismic data of COROT [32] and KEPLER [33]. TOPS spectra database from Los Alamos [1, 28] is also available on line. One difficulty for designing an experiment using on line results is that they are tabulated on temperature-density grids so that interpolations are necessary for intermediate cases. The FLYCHK code [29] can also provide detailed spectra for any thermodynamic conditions, but its on line version is not appropriate for low density stellar conditions.

Recently the opacity codes (OPAS [27], SCO [24] and SCO-RCG [30]) have been developed and applied to some specific astrophysical conditions. These codes contain different levels of modeling and approximations, from self-consistent average atom to detailed level accounting. However, the results of these new codes are not available on line.

We cover the theoretical results of the consortium OPAC for iron and nickel opacities in the rho-T phase space region where iron and nickel play an important role in forming the Rosseland mean, which is about 20 eV and $\rho \sim 3 \cdot 10^{-3}$ g/cm³ for all the iron-group lines. Comparisons between opacity calculations will be presented here for several conditions found in the LULI 2010 experiments [22]. The spectral range of the transmission experimental data is between 60eV and 160 eV. Even though large differences appear in these comparisons, which may lead to opacity code modifications, these comparisons are instructive for the atomic physics and plasma community.

### 2.1 Atomic Opacity codes

First, we compare the mean ionization states, the Rosseland and Planck mean opacities. Further, for select cases, we show the distribution of ionic populations and the contributions of the bound-bound (bb), bound-free (bf), free-free (ff) opacities to the total opacity contribution.

The eight opacity codes, namely OPAS, OP, SCO, SCO-RCG, LEDCOP, STA, Cassandra, and FLYCHK codes have already been documented in [5]. Here we give brief comments describing their contributions to this work.

The « Opacity Project (OP)» is an on line atomic database [3] where data are computed in the close-coupling approximation by means of the R-matrix method including tens or a few hundreds of configurations. The main efforts of OP team have been focused on iron. However, with the use of a scaling method data for numerous other elements, including nickel, are available. From the OP data one can produce frequency-dependent opacities and ionic population distributions for the tabulated temperature and density values [3]. The values shown in this paper are directly extracted from these tables.

The local thermodynamic equilibrium (LTE) OPAS detailed radiative opacity code has been recently applied to solar conditions [18]. Total absorption spectra and ionic population distributions have been calculated for the 30 elements generally used in astrophysics and specific calculations have been performed for the present studies.

The recently updated version of the STA opacity code interpolates smoothly between the average-atom results and the detailed configuration accounting that underlies the unresolved transition array (UTA) method. Thus, computing time is substantially reduced. Results often look like smoothed detailed calculations – see for example Fig. 4a, b and Fig. 6 of Sec. 3. The modified STA code can be used to check density or/and temperature effects and all the component contributions are available.

SCO-RCG is a LTE hybrid opacity code that combines the statistical Super-Transition-Array approach and fine-structure calculations for intense and spectrally broad transition arrays [30]. The SCO-RCG code was designed for the diagnosis and quantitative interpretation of spectroscopy experiments in LTE plasmas. It ensures thermodynamic consistency, requires limited resources, and is robust enough to generate opacity tables. The statistical part of the spectrum is supplied by the super-configuration code SCO (Super-Configuration Opacity) [24]. The Slater, spin-orbit and dipolar integrals are calculated by SCO, that provide an accurate description of the plasma screening effects on the wave-functions. They are then passed to an adapted RCG routine of R. D. Cowan [34]. This RCG routine calculates the level energies and E1 transitions. SCO-RCG takes into account the electrostatic interaction between parent relativistic configurations, i.e., relativistic configurations belonging to the same non-relativistic configuration. The resultant total opacities and populations are available.

The web-accessible FLYCHK population kinetics code [29], generates atomic populations and charge state distributions. It is used here even though the on-line version is not precise enough for the low density conditions of our applications and for the spectral range where $\Delta n=0$ transitions contribute significantly. The poor representation for $\Delta n=0$ transitions arises because the code uses a hydrogenic approximation that performs poorly for $\Delta n=0$ transitions. Populations and spectra together with the detail of the bb, bf, and ff contributions are given.

CASSANDRA's self-consistent field calculation LTE opacity code [25] gives absorption contributions and mean ionization.

Results from the The Light Element Detailed Configuration OPacity (LEDCOP) code for iron and nickel are also shown [28]. In LEDCOP each ion stage is considered in detail and interactions with the plasma are treated as a perturbation.

## 2.2 Selection of the Mean Ionization

The density of stellar envelopes is sufficiently small that the temperature and density conditions cannot be reproduced in the laboratory. Therefore, to test the codes in the regimes that provide insight into the opacity of stellar envelopes we consider some conditions that lead to similar mean ionization states [5, 20] to reproduce comparable ionization distributions.

The importance of the iron opacity for temperatures near 27.3 eV and densities near 3.4 g/cm$^3$, which corresponds to a mean ionization, <Z> ~ 9 has been discussed elsewhere [5, 20]. In these conditions one can check the different ionic contribution to the iron peak in β Cephei and for this reason has been chosen for the LULI 2010 experiment. We also compared opacities at the neighboring temperatures of 15.3eV and 38.5eV to estimate the effects of gradients and time evolution in the experiment. Table 1 displays the six parameters of material, density, and temperature defined for the theoretical comparisons. These conditions yield the same electron density of $N_e =3.16\ 10^{20}$ cm$^{-3}$. The corresponding mass densities calculated with OP for the <Z> are given in Table 1.

| Case | A | kT ( eV) | ρ (mg/cm³) | <Z>OP |
|---|---|---|---|---|
| 1 | Fe | 15.3 | 5.48 | 5.35 |
| 2 | Fe | 27.3 | 3.39 | 8.65 |
| 3 | Fe | 38.5 | 2.63 | 11.2 |
| 4 | Ni | 15.3 | 5.65 | 5.46 |
| 5 | Ni | 27.3 | 3.67 | 8.39 |
| 6 | Ni | 38.5 | 2.73 | 11.3 |

Table 1. Thermodynamic conditions of the six test cases.

The relationship between electron and mass density has been applied to obtain the input density and temperature conditions for the opacity codes. We have checked with FLYCHK and STA codes that this choice has no influence on the results for the analyzed conditions. For these conditions LTE is assumed.

In Fig. 1 we compare the mean ionization stages obtained by the different codes to assess the influence of this choice of <Z> on our comparisons. Results are displayed for Fe and Ni, and for the 3 pairs of temperature and density. The results for two LTE average atom codes have been added for the comparison of the mean ionization stage. The free electron definition of the AA_$Z_P$ code describes the density of electrons with positive energy in a quantum description, see definition 2 in Ref. [35], and the free electron definition of the AA_$Z_M$ code is the definition of More [36] which involves only the electronic density at the edge of the box.

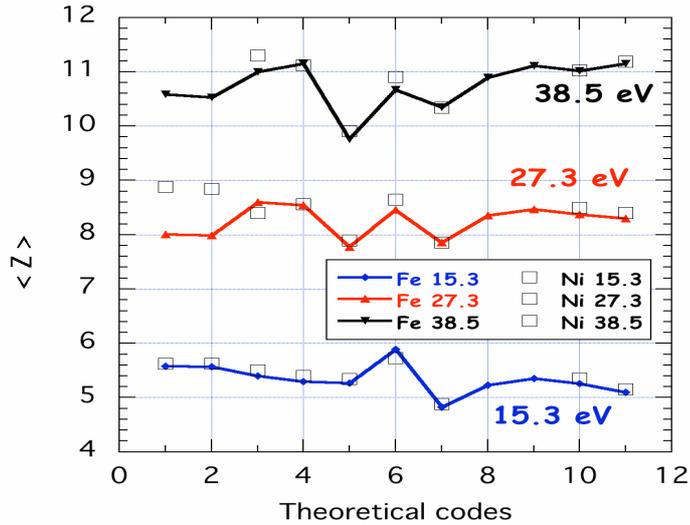

Fig. 1. Iron (solid line) and nickel (□) mean ionization states given by eleven codes for the 6 cases defined in Table 1 and corresponding to the 3 temperatures (15.3, 27.3 and 38.5eV). From left to right the codes used are: FLYCHK (NLTE):1, FLYCHK (LTE):2, OP:3, STA:4, AA_Z_P:5, AA_Z_M:6, CASSANDRA:7, OPAS:8, SCO(relativistic):9, SCO-RCG:10, LEDCOP:11. Codes 5 and 6 are LTE Average Atom Ionization models [34-35].

Differences in mean ionization state obviously imply differences in ionic population distributions and suggest differences in frequency-dependent opacity for each thermodynamic condition.

Fig. 2a shows a comparison between OPAS, OP, STA, SCO and FLYCHK for the iron ionic population distributions at the temperature of case 1, 15.3 eV. Fig. 2b, c show the same comparisons for the mean iron conditions for the temperatures of cases 2 and 3, 27.3 eV and 38.5 eV, respectively. In addtion, the HULLAC ionic population results for the main ion contributions have been added for the case 2 and case 3.

We note that different models do not show exactly the same results.

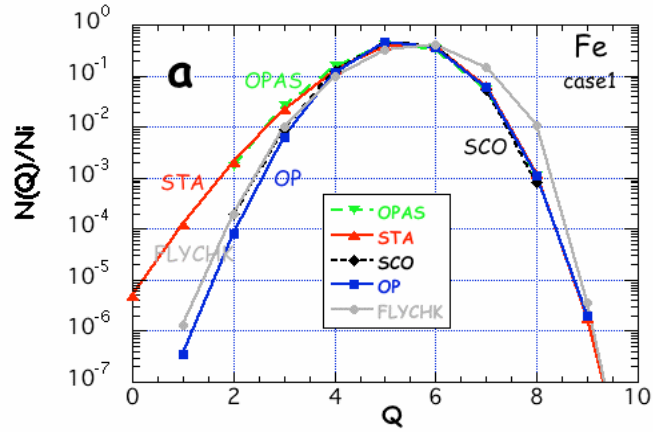

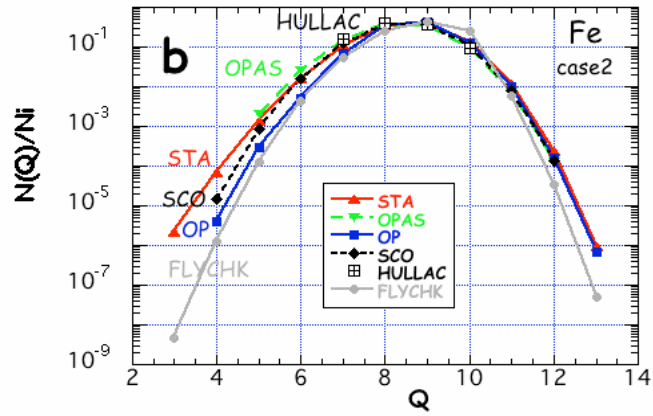

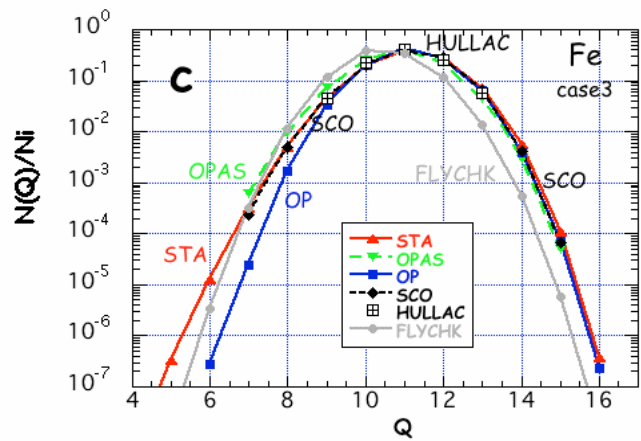

Fig. 2a, 2b, 2c. Comparison between iron ionic population distributions of theoretical codes for conditions of case 1 to 3 of Table 1. a): case1, OPAS (green long-dashed line, STA (red solid line + ▲), SCO (black dashed line), OP (blue solid line + ■), FLYCHK (gray solid line + ●) codes; b): case2, OPAS, STA, SCO, OP, FLYCHK, HULLAC (⊞) codes; c): case3, OPAS, STA, SCO, OP, FLYCHK, HULLAC (⊞) codes.

## 3. Photon energy spectra of iron and nickel

In support of the LULI campaign of measurements of the photo-absorption spectra of different medium Z plasmas, we focus on temperatures between 15 and 40 eV and densities of 2 to 5 mg/cm$^3$. Iron and nickel frequency-dependent opacity spectra predictions are shown associated with comparisons of the corresponding Rosseland and Planck mean opacity values obtained by the eight theoretical calculations. Even if the accessible spectral range of the spectrograph for the experimental results of the year 2010 [22] is between 60eV and 160 eV, we shall often compare the codes on a larger interval, up to 250eV.

### 3.1 Iron theoretical opacity calculations

Theoretical iron opacity spectra for case 2, i.e., at T = 27.3 eV have been shown in Ref. [5], so here we display in Fig. 3a, b the two other temperatures (15.3 and 38.5 eV). Between the three cases the variation in temperature, ±40%, is much larger than the variation of density and its influence on the spectra is more important.

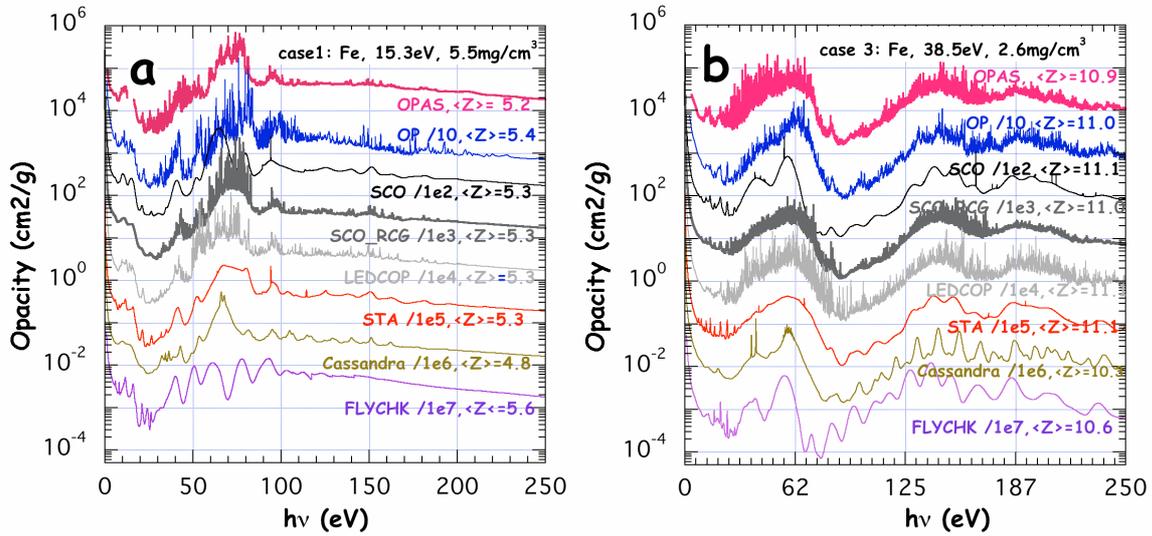

Fig. 3a, 3b. a) Comparison of 8 iron theoretical monochromatic spectra for T= 15.3 eV and ρ= 5.5 10$^{-3}$ g/cm³ (case 1 of Table 1). Each spectrum is shifted by a decade for clarity. From top to bottom OPAS, OP, SCO, SCO-RCG, LEDCOP, STA, CASSANDRA and FLYCHK code results. b) Same comparison for T= 38.5eV and ρ= 2.6 10$^{-3}$ g/cm (case 3 of Table 1). Case 2 of Table 1 has been already published, see Fig. 4 of Ref. [5]. The mean ionization charge follows the name of the codes.

As for case 2 in Ref. [5], the iron spectra of cases 1 and 3 clearly show differences between results.

It is difficult to present a detailed analysis of these results because the ionic populations are different from code to code for each case and so, therefore, are the mean ionization values. Nevertheless, it is possible to distinguish the 3 families of results: statistical (SCO, CASSANDRA, STA), detailed (OPAS, LEDCOP, OP) or mixed (SCO-RCG). The FLYCHK code clearly show difficulty in reproducing the details of the theoretical spectrum, as given by most of the codes. The statistical code SCO and the Average Atom CASSANDRA give shifted spectra with little detail. Even if the STA code is considered as a statistical code its level of description is much more sophisticated and results appear, in this range of energy values, as a smooth fit of the detailed OPAS or LEDCOP spectra (Fig. 4a). The OP results show definitive differences with the OPAS, LEDCOP and SCO-RCG spectra (Fig. 4b). These last 3 detailed codes, OPAS, LEDCOP and SCO-RCG, give very similar results even if, as expected, the level of description of the lines is not exactly the same. We recall that our conclusions are only valid for the spectral range and thermodynamic conditions discussed in this paper.

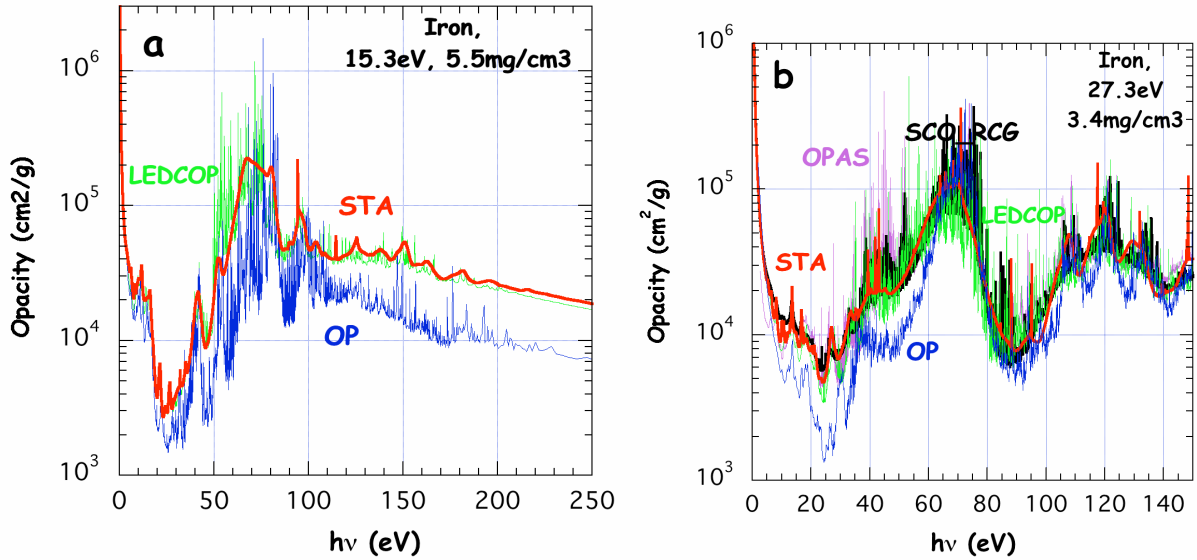

Fig. 4a, 4b. a) Comparison of the theoretical LEDCOP (green line), OP (blue line), STA (red line) iron monochromatic opacity spectra versus energy. The mean conditions used for the theoretical calculations were those of case 1 (15.3eV, $\rho$= 5.5 $10^{-3}$ g/cm$^3$). b) Comparison of the theoretical LEDCOP (green line), OP (blue line), STA (red line), OPAS (purple line) and SCO-RCG (black line) iron monochromatic opacity spectra versus energy. The mean conditions used for the theoretical calculations were those of case 2 (27.3eV, $\rho$= 3.4 $10^{-3}$ g/cm$^3$).

## 3.2 Nickel opacity calculations

The same comparisons have been repeated for nickel for conditions of cases 4, 5 and 6, using the OP, SCO-RCG, STA, CASSANDRA and FLYCHK codes. In Fig. 5a, 5b, 5c we show the frequency-dependent opacities and in Fig 5d, e, f we show the abundances. Other comparisons have been made between the eight codes to confirm the preceding conclusions on iron. As was said before, the level of description in the on line version of FLYCHK is not show because it lacks spectral precision. CASSANDRA results are slightly shifted in energy and the STA spectrum acts as a fit of the SCO-RCG detailed description. OPAS, SCO-RCG, LEDCOP and STA are in good agreement for similar nickel conditions.

We can observe that a detailed description of the transitions appears for higher temperatures in the OP nickel spectra as compared to iron spectra (Fig. 5a). Thus, differences are emphasized for the low temperature case between OP and other codes in case of nickel, seven though the corresponding ionic populations do not show so large differences. This is illustrated in Fig. 6, where LEDCOP, SCO-RCG, STA and OP codes are compared.

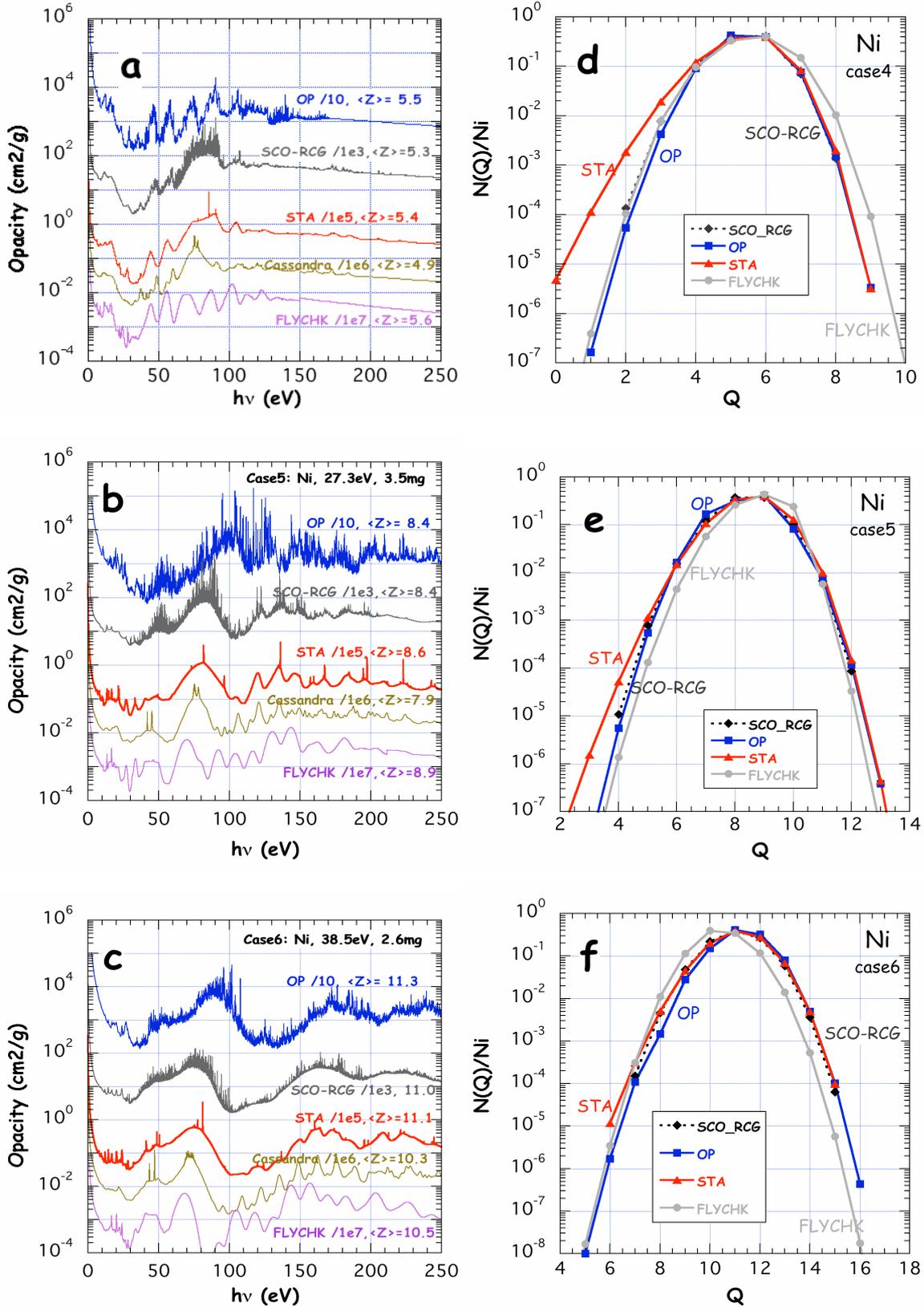

Fig. 5a, 5b, 5c, 5d, 5e, 5f. a), b), c) Comparaisons of five theoretical monochromatic nickel opacity spectra versus energy. From top to bottom OP (blue line), SCO-RCG (black line), STA (red line), CASSANDRA (green line), FLYCHK (purple line) : (a, case 4), (b, case 5), (c, case 6). The mean ionization charge follows the name of the codes. Vertical shifts between spectra follow shifts used in Fig.3. d), e), f) Comparison between nickel ionic population distributions of theoretical codes for the same conditions corresponding to cases 4 to 6 of Table 1: STA (red solid line + ▲), SCO-RCG (black dashed line), OP (blue solid line + ■), FLYCHK (gray solid line + ●).

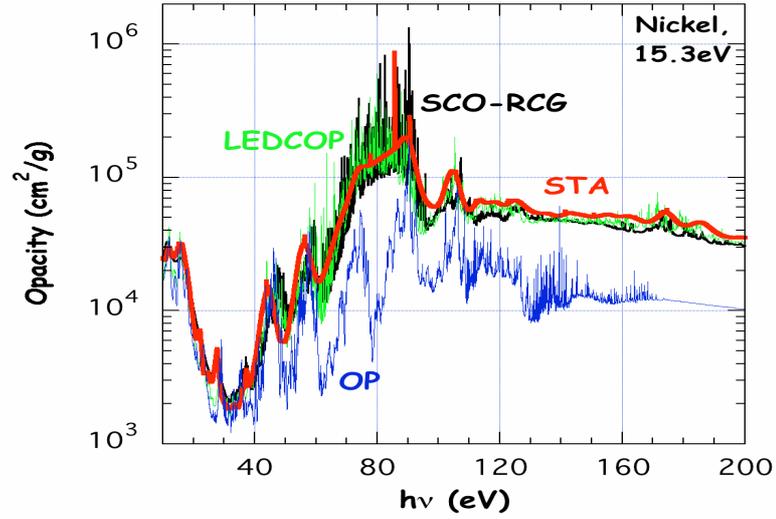

Fig. 6. Comparison of the theoretical LEDCOP (green line), OP (blue line), STA (red line) and SCO-RCG (black line) nickel monochromatic opacity spectra versus energy. The mean conditions used for the theoretical calculations were those of case 4 (15.3eV, $\rho$= 5.65 $10^{-3}$ g/cm$^3$) which are very near to the ones obtained by hydro-simulation performed for shot 42 of the LULI 2010 experiment [22, 23].

Moreover some comparisons of the bound-bound (bb) and bound-free (bf) contributions of the frequency-dependent opacity are presented for the STA and LEDCOP results in Fig. 7a. Two nickel conditions are presented T=15 eV, $\rho$=2x$10^{-3}$ g/cm$^3$ and T=36 eV, $\rho$=4 $10^{-3}$ g/cm$^3$ and show. These comparisons show good agreement between results. The main contributions come from bb contributions for energies below 100 eV. The results for the STA total, bb, bf, ff and total contributions to the opacity are shown in Fig. 7b.

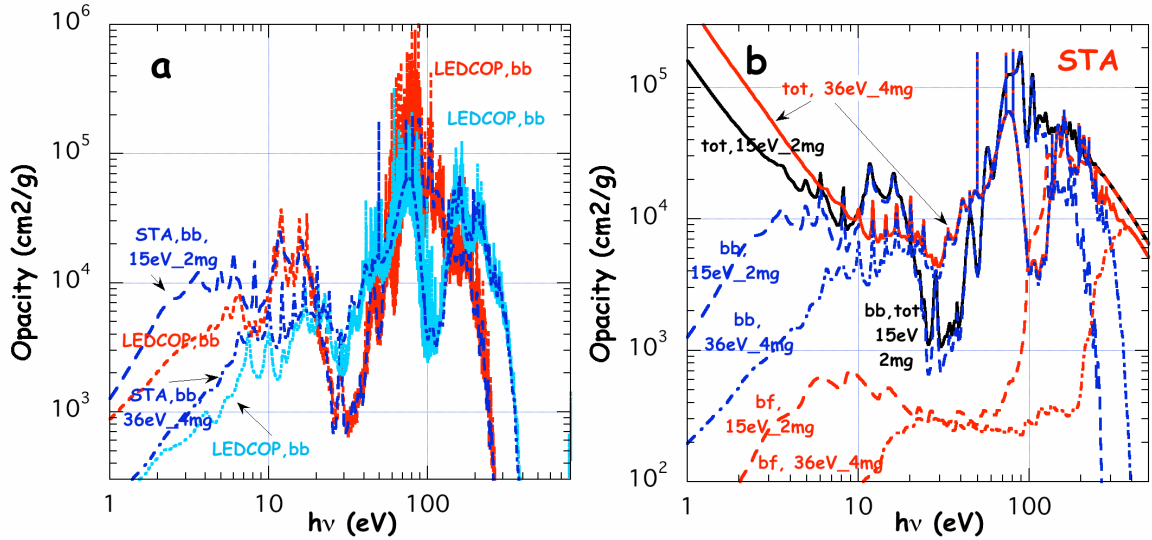

Fig. 7a, 7b. a) Comparison of STA (long-dashed and dotted-dashed dark blue lines) and LEDCOP (dashed red and dotted sky blue lines) nickel bound-bound (bb) contributions of the monochromatic opacity versus energy for T= 15eV, $\rho$= 2 $10^{-3}$ g/cm$^3$ and T= 36eV, $\rho$= 4 $10^{-3}$ g/cm$^3$. b) STA bound-bound (bb, long-dashed and dotted-dashed blue lines), bound-free (bf, long-dashed and dotted-dashed red lines) and total (tot, black and red solid lines) contributions of the monochromatic opacity for the same thermodynamic conditions as Fig. 7a.

## 3.3 Temperature effects

The determination of the temperature, more so than the density, is very important for the comparison with the experimental data. Figure 8 shows the sensitivity of the calculated STA opacities with respect to the temperature. For clarity we do not superimpose results from other codes (LEDCOP, OPAS, SCO-RCG and CASSANDRA), but the variations are similar.

It is clear that a 30% change in temperature around T=15eV has a large effect on the shape of the spectra and that temperature can be used as a diagnostic of the LULI experiment data.

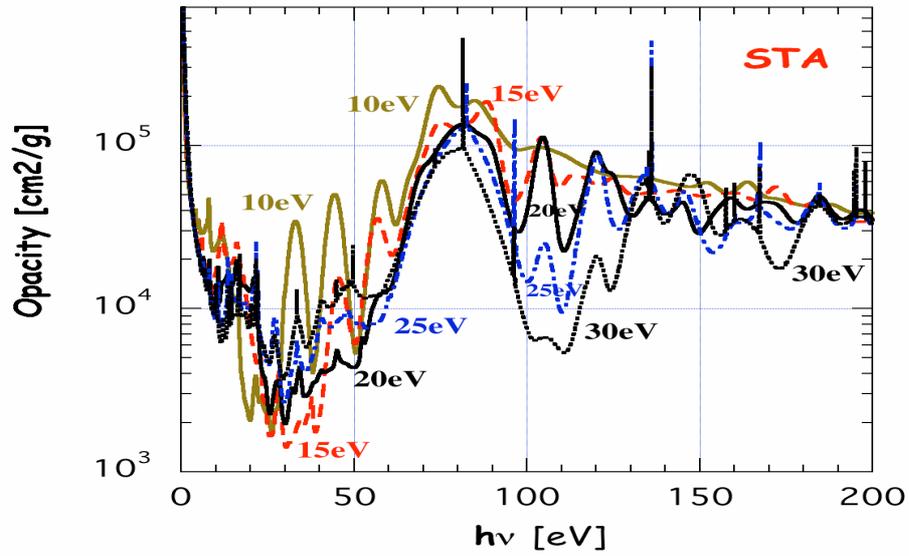

Fig. 8. Temperature influence on the STA total monochromatic opacities versus energy for ρ= 3.4 mg/cm3. The displayed temperatures: 10, 15, 20, 25 and 30eV are recalled on the plot. The large structures disappear at higher temperatures.

## 3.4 Rosseland and Planck means

In LTE the total absorption opacity coefficient $\rho\kappa_{tot}$ to a photon with energy $h\upsilon$ is the sum of the true absorption $\rho\kappa_{tot}^a$, corresponding to the absorption of the bb ($\mu_{bb}$), bf ($\mu_{bf}$) and ff ($\mu_{ff}$) transitions including the correction for induced emission, and of the scattering contribution from free electrons $\mu_{sc}$,

$$\rho\kappa_{tot} = \rho\kappa_{tot}^a + \kappa_{sc} = (\mu_{bb} + h\mu_{bf} + \mu_{ff})(1 - e^{-h\upsilon/kT}) + \mu_{sc}.$$

$\rho$ is the mass density, T the temperature, k the Boltzmann constant.

The corresponding Rosseland and Planck opacity means, respectively $K_R$ and $K_P$, are given by the well-known expressions,

$$\frac{1}{K_R} = \frac{\int d\upsilon \, \frac{1}{\kappa_{tot}} \frac{dB_\upsilon}{dT}}{\int d\upsilon \, \frac{dB_\upsilon}{dT}}, \quad K_P = \frac{\int d\upsilon \, \kappa_{tot}^a B_\upsilon}{\int d\upsilon \, B_\upsilon},$$

where $B_\upsilon$ is the spectral Planck function. Scattering contributions $\mu_{sc}$ have been neglected in the Rosseland mean values presented here. Thus, with $u = h\nu/kT$, $K_R$ and $K_P$ are given by,

$$\frac{1}{K_R} = \frac{15}{4\pi^4} \int_0^\infty \frac{1}{\kappa_{tot}^a} \frac{u^4 e^{-u}}{(1-e^{-u})^2} du,$$

$$K_P = \frac{15}{\pi^4} \int_0^\infty \kappa_{tot}^a \frac{u^3 e^{-u}}{(1-e^{-u})} du,$$

The results are illustrated in Fig. 9a, 9b for the iron cases and in Fig. 10a, 10b for the nickel cases. As expected all results are in broad agreement for the highest temperature but differences appear at low temperature and correspond to the differences between the opacity spectra.

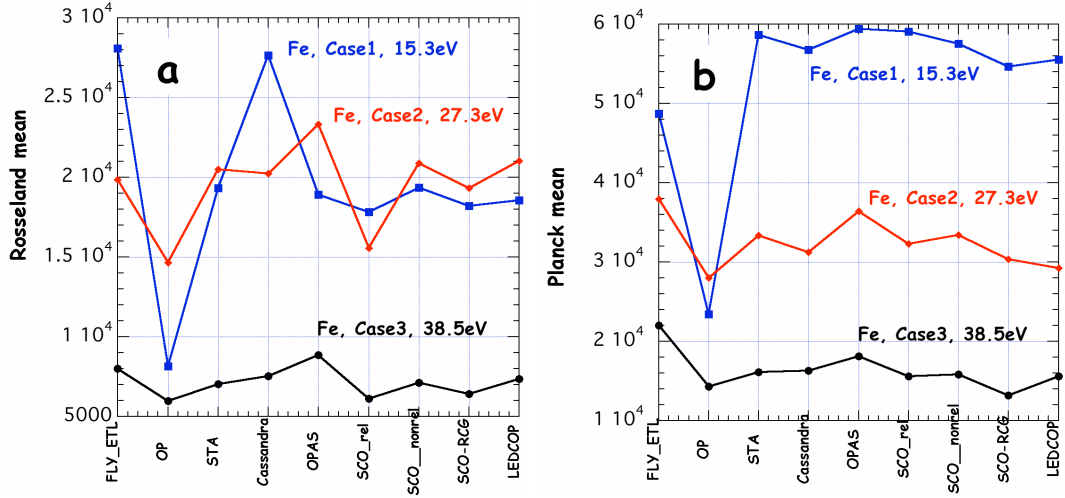

Fig. 9a, 9b. Comparisons of Iron Rosseland (a) and Planck (b) means for nine codes for mean conditions corresponding to first three cases of Table 1.

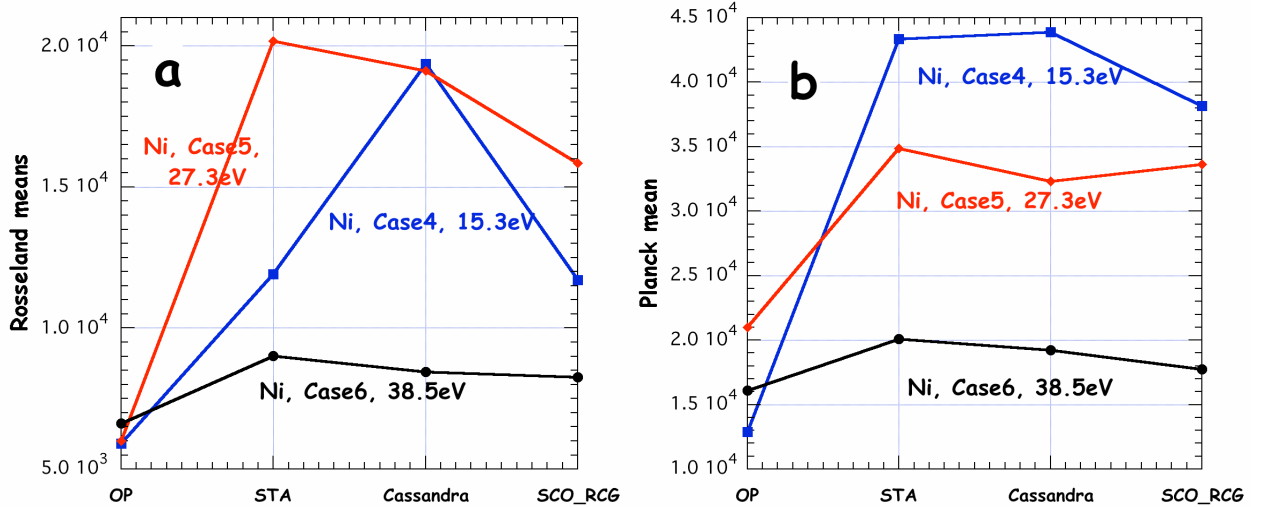

Fig. 10a, 10b. Comparisons of nickel Rosseland (a) and Planck (b) means for four codes for nickel mean conditions of cases 4, 5 and 6 of Table 1.

## 4. Conclusion

We have compared a large number of opacity calculations for experimental conditions relevant to stellar envelop conditions of intermediate masses. More comparisons are in progress including frequency-dependent spectra from HULLAC. The codes are based on different approaches for the description of levels and lines: statistical (SCO, STA), detailed (OPAS, OP, FLY, HULLAC, LEDCOP) or mixed (SCO-RCG). In the case of OP and HULLAC, full configuration interaction (CI) is included. However, the SCO-RCG and STA codes partially take this effect into account, including interaction allowed only between levels of the same non-relativistic configuration. These comparisons show important differences in the frequency-dependent spectra, with a distinct behavior for the OP results. Differences decrease when the temperature increases as one can easily imagine. The SCO, CASSANDRA, FLYCHK and STA code results present expected differences among them, mainly less structure or shifts in energies, due to their simplified level of atomic description. However the OPAS, SCO-RCG and STA results remain comparable. Some differences between the detailed theoretical spectra are related to the difference in the ionization distribution, with the influence of the contribution of the tail of the distribution being observational. The most important feature of these comparisons is whether the codes include CI or not, and how many configurations are included. This is related to the difficulty of taking into account numerous excited levels in the R-Matrix formalism. We must point out that OP calculations for nickel are not purely ab initio but are derived from iron data. New OP calculations are in progress as well as HULLAC calculations to better interpret the observed differences. As expected, the temperature dependence of frequency-dependent opacity is more important than the density dependence for the conditions of the experiment.